# Sub2Full: split spectrum to boost OCT despeckling without clean data

Lingyun Wang,[1,2] Jose A Sahel,[1] and Shaohua Pi[1,2,*]

[1]Department of Ophthalmology, University of Pittsburgh, Pittsburgh, PA 15213, USA
[2]Department of Bioengineering, University of Pittsburgh, Pittsburgh, PA 15213, USA

*shaohua@pitt.edu

**Optical coherence tomography (OCT) suffers from speckle noise, causing the deterioration of image quality, especially in high-resolution modalities like visible light OCT (vis-OCT). The potential of conventional supervised deep learning denoising methods is limited by the difficulty of obtaining clean data. Here, we proposed an innovative self-supervised strategy called Sub2Full (S2F) for OCT despeckling without clean data. This approach works by acquiring two repeated B-scans, splitting the spectrum of the first repeat as a low-resolution input, and utilizing the full spectrum of the second repeat as the high-resolution target. The proposed method was validated on vis-OCT retinal images visualizing sublaminar structures in outer retina and demonstrated superior performance over conventional Noise2Noise and Noise2Void schemes. The code is available at https://github.com/PittOCT/Sub2Full-OCT-Denoising.**

http://doi.org/10.1364/OL.XXXXXX

Optical coherence tomography (OCT) is a reflectance-based, depth-resolved, micron-resolution imaging technique that has become a routine imaging modality in ophthalmology [1]. However, the speckle noise, induced by the interference of light, obscures the implementation of ideal axial resolution to visualize the fine details of bio-tissue, bringing challenges to the subsequent image analysis and interpretation [2].

Conventional OCT denoising approaches typically rely on sparse reconstruction or frame averaging (usually >100 frames) [3-5]. The major drawback of sparse reconstruction methods is the potential blurry of image, sacrificing the attainable resolution of OCT. Additionally, the frame averaging scheme subjects to long acquisition time (especially for volume averaging) and precise image registration, making it susceptible to eye movements and less practical in clinical applications [6].

With the rapid advancement of deep learning methods, various networks, including both the supervised and self-/unsupervised strategies, have been proposed for OCT despeckling. For instance, Ma et al. introduced a supervised cGAN method relying on frame-merged noise-free images for training [7]. Although effective, the supervised methods pose a shared challenge of obtaining noise-free images. Recently, the self-/unsupervised methods are emerging to address this problem, requiring only noisy images for the training and processing. Lehtinen et al. proposed a Noise2Noise (N2N) approach to restore images by using two independent observations of the same scene to estimate the real structures [8]. Additionally, blind-spot schemes such as Noise2Void (N2V) used only one observation per scene by strategically masking the central pixel of a patch during training, treating the masked central pixel as the target for the network [9, 10]. However, as gradient information is backpropagated only for the masked pixels, N2V suffers a compromised enhancement to N2N. Nevertheless, performance of self-/unsupervised methods is usually less optimal to supervised ones.

Visible light OCT (vis-OCT) is an emerging modality with higher scattering contrast and improved axial resolution than standard near-infrared OCT (NIR-OCT) [11]. These benefits enable the visualization of retinal fine structures in inner plexiform layer and outer retinal bands [12, 13]. However, the strong speckle noise undermines these advantages, making the challenging frame averaging process an essential procedure for vis-OCT. Fortunetaly, attempts are emerging to develop deep learning despeckling method for vis-OCT. Notably, Fan et al. employed N2N strategy to reduce noise in Fibergrams [14]. Ye et al. achieved denoising in the proposed DenoiSegOCT framework using N2V [15].

Recognizing the need for effective and robust self-supervised OCT denoising solutions without clean data, we proposed an innovative Sub2Full (S2F) strategy in this study. Building upon the strengths of N2N (we choose N2N here as OCT routinely acquires at least two repeated B-scans for applications like OCT angiography), it works by generating a low-resolution (LR) image for expanded mapping to multiple underlaying high-resolution (HR) images, and thus increases the possibility to converge to the true clean image. The S2F utilizes the sub-band image of the first repeated B-scan (S-R1) as the LR input, while setting the full spectrum image of the second repeated B-scan (F-R2) as the HR target for the network. We showed that the network powered with S2F strategy achieved superior performance over conventional N2N and N2V schemes, allowing identification of sublaminar structure of outer retina in single B-scan frame without averaged clean data in vis-OCT. Although validated in vis-OCT with N2N, it should be noted that S2F strategy can potentially be applied to all OCT types and integrated with various schemes for improved denoising performance.

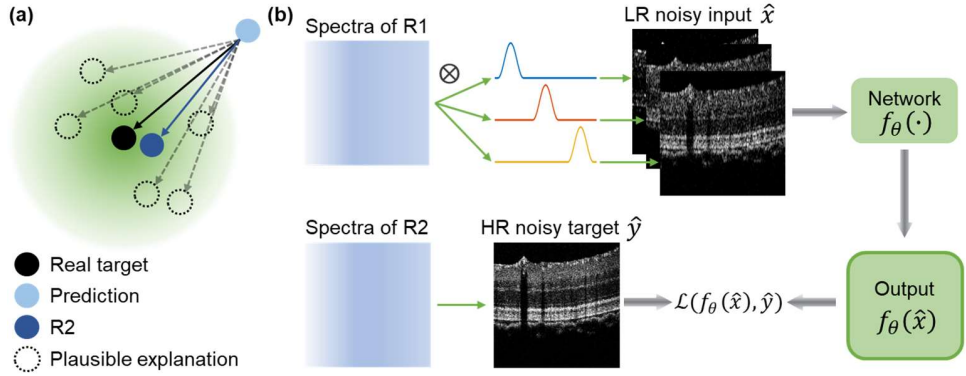

Fig. 1. (a) Schematic diagram of super-resolution denoising effect. (b) Framework of the proposed Sub2Full denoising scheme. The Gaussian window was used to split the spectrum and generate low-resolution sub-band images.

To delve into the principles of the proposed S2F scheme, it is essential to understand the mechanism of traditional supervised and self-/unsupervised deep learning denoising methods. Considering speckle noise, reflectance of OCT B-scan image can be expressed by $y_1 = x + n_1$ where clean image $x$ is corrupted by a zero-mean independent speckle noise $n_1$. Therefore, the endeavor of supervised denoising method is to minimize the discrepancy between the noisy input $y_1$ and the corresponding clean target $x$ through Eq. (1):

$$\mathbb{E}[\| f_\theta(y_1) - x \|_2^2] = \mathbb{E}[\| f_\theta(y_1) \|_2^2 - 2x^T f_\theta(y_1) + \| x \|_2^2], \quad (1)$$

where $\|\cdot\|_2^2$ denotes the L2 norm, and $\theta$ is the trainable parameters of a network $f$, and $\mathbb{E}$ denotes the expectation. Since $\mathbb{E}[\| x \|_2^2]$ is a constant, the network training becomes the searching process for an optimal solution of $\theta$ using Eq. (2):

$$\arg\min \mathbb{E}[\| f_\theta(y_1) \|_2^2 - 2x^T f_\theta(y_1)]. \quad (2)$$

For N2N method, there are two repeated B-scans acquired from the object and the second repeated image can be expressed as $y_2 = x + n_2$. Similar to Eq. (1), the two independent observations (B-scan image) serve as noisy input and corresponding noisy target in Eq. (3), respectively:

$$\begin{aligned}\mathbb{E}[\| f_\theta(y_1) - y_2 \|_2^2] &= \mathbb{E}[\| f_\theta(y_1) - x - n_2 \|_2^2] \\ &= \mathbb{E}[\| f_\theta(y_1) - x \|_2^2 + \| n_2 \|_2^2 \\ &\quad - 2n_2^T(f_\theta(y_1) - x)] \\ &= \mathbb{E}[\| f_\theta(y_1) \|_2^2 - 2x^T f_\theta(y_1) + \| n_2 \|_2^2 \\ &\quad + 2n_2^T x] - 2\mathbb{E}[n_2^T]\mathbb{E}[f_\theta(y_1)],\end{aligned} \quad (3)$$

where $\mathbb{E}[\| n_2 \|_2^2 + 2n_2^T x]$ is a constant scalar and $\mathbb{E}[n_2^T]$ is zero as the noise is zero-mean. Hence, the optimization solution of $\theta$ in N2N is determined by $\arg\min \mathbb{E}[\| f_\theta(y_1) \|_2^2 - 2x^T f_\theta(y_1)]$, similar to the supervised method with clean image. As long as the noise is zero-mean and independent, optimization of N2N can ideally converge to the same solution of supervised training even if the noise distribution in two observations changes. However in reality, N2N cannot achieve results comparable to supervised methods unless trained on an infinitely large dataset [8]. As illustrated in Fig. 1(a), when the resolution of the input R1 and the target R2 is similar, the gradient tends to point the target (denoted by a blue arrow) with a 1:1 mapping, while a single observation of the object cannot encapsulate the real scene (real target).

Inspired by the fact of unexpected denoising effect [16,17] in OCT super-resolution studies, and therefore output fuzzy images [18], we realized that a single low-resolution (LR) image might have multiple plausible explanations for the high-resolution (HR) image. Here, we proposed the Sub2Full strategy by employing LR images obtained from splitting the spectrum of R1 while using high-resolution (HR) images of R2 as the target, taking advantage of the 1:N mapping for searching the optimal solution in denoising methods. As shown in Fig. 1(a), the 1:N mapping will force the gradient to propagate in the direction of each plausible explanation (denoted by gray dashed arrows), ultimately achieving improved optimization of pointing to the real target. The framework of the Sub2Full strategy is summarized in Fig. 1(b) and described as below, i) acquiring two repeated B-scans from the same object, ii) reducing the resolution of R1 by multiplying the spectrum with a Gaussian window, iii) feeding the resulted S-R1 images into the network, with the target set as the full spectrum images of second repeat F-R2, iv) using the L2 norm, which is typically used to suppress zero-mean speckle noise, to minimize the distance between the prediction and the target. It should be noted that for each B-scan, more than one sub-band can be obtained by varying the center wavelength of Gaussian window and be incorporated in different iterations of training process for improved spatial perspectives and boarder range of information.

To evaluate the performance of S2F scheme, vis-OCT retinal scans were acquired in brown Norway rats using a custom-built prototype [6]. The system has a full-width half-maximum bandwidth of 90 nm from 510 to 610 nm, operated at a 50 kHz A-line sampling rate. Each volume contains 500 A-lines per B-scan, 2 repeated frames for each B-scan, and 500 B-scans in total. The interferogram of each scan was recorded by a line scan camera (Basler spl4096-140km) and further processed in MATLAB to resolve the OCT images with split spectrum or full spectrum. Ethics approval for the protocols was obtained from the Institutional Animal Care and Use Committee (IACUC) of the University of Pittsburgh.

To make a fair comparison, classic U-Net was used as the backbone architecture for all models in this paper [8]. For training and validation, 1200 B-scans from four OCT volumes were selected. The other (n=15) OCT volumes acquired at different retinal regions and animals were used for testing. In S2F, the 1200 noisy input S-R1 images were generated from 400 B-scans in the first repeat with spectral windows of 50% bandwidth of the full spectrum centered at short, medium, and long wavelength. The corresponding F-R2 images were employed as the noisy target. To avoid over-fitting, we divided the dataset with a ratio of 4:1 for network optimization and validation. The experiments were implemented in Python 3.7 with Pytorch 1.13.1 platforms using INTEL i9-13900K CPU and NVIDA RTX 3080 Ti GPU. We used a batch size of 2 and Adam optimizer for training with an initial learning rate of 1e-3. All models were trained by 200 epochs to ensure convergence.

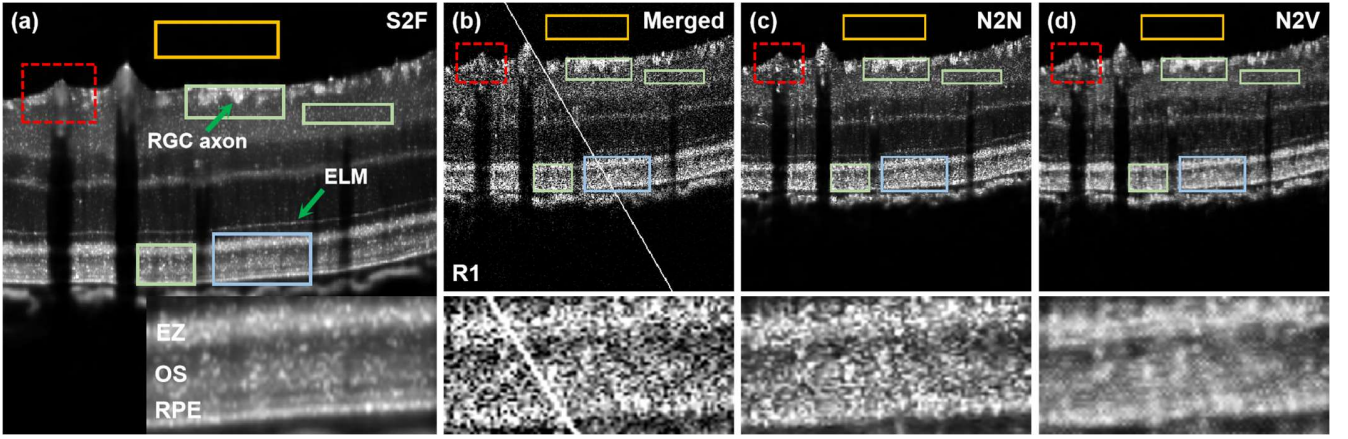

Fig. 2. Comparison of different methods in despeckling rat retinal vis-OCT images. (a) The results of S2F scheme. (b) The left part is the single B-scan image, and the other is the merged (N=2) B-scan image. (c)-(d) The results of N2N and N2V scheme, respectively. The zoomed region framed with a light blue rectangle in each B-scan image is shown below. One background region framed with an orange rectangle and three structure regions with light green rectangles were selected for metric calculation. RGC axon, retinal ganglion cell axon bundles; ELM, external limiting membrane; EZ, ellipsoid zone; OS, outer segment; RPE, retinal pigment epithelium.

Fig. 2(a) showed an enhanced B-scan by S2F. In contrast to the raw and merged B-scans in Fig. 2(b), the strong speckle noise was significantly suppressed by S2F, allowing better distinction of blood vessel and retinal ganglion cell (RGC) axon bundles from neighboring retinal tissue. Through visual inspection, the S2F demonstrated superior performance than the N2N in Fig. 2(c) and N2V in Fig. 2(d) that the image by N2N exhibited excessive granny appearance in outer retina while that by N2V had checkerboard artifacts attributed to blind-spot scheme. The visibility of retinal layers and boundaries, especially the external limiting membrane (ELM), was greatly improved in S2F. More importantly, multiple sublaminar structures in outer retina, or more specifically, the outer segment (OS) layer and the RPE layer, were identified, which is unique for S2F when compared to other models.

To clarify the identified outer retinal sublaminar structures, we corelated the enhanced vis-OCT images (Fig. 3(a)) to a diagram of the outer retina (Fig. 3(b), reproduced from [19] with permission), as well as transmission electron microscopy (TEM) images obtained from the same species (Fig. 3(c), reproduced from [20] with permission). In vis-OCT image, a noticeable gradient increase of reflectance from basal to distal was observed for the OS, which should be attributed to the continuous renewal process of photoreceptor outer segments [21]. Corresponding to the diagram and TEM images, the bright sub-layer positioned adjacent and beneath to the OS should be the RPE microvilli (mv). The following dark sub-layer should be the RPE melanin (ml) granules as its weak reflectance could be explained by the strong light absorption of the melanin to protect the cell damage from photosensitized oxidations by absorbing excessive light. The two bottom layers represented the RPE mitochondria (mc) and basal infoldings (bi), respectively. The good correlation between vis-OCT and TEM, as well as existing knowledge, indicated that restoration achieved by S2F not only improved the general appearance of the retinal images but also resolved critical sublaminar features that can only be appreciated with TEM, providing a comprehensive and faithful representation of the retinal structures.

Next, to quantitatively evaluate the denoising performance, three image metrics including signal-to-noise ratio (SNR), contrast-to-noise ratio (CNR), and variance (VAR) were employed:

$$SNR = 20\log(I_{max}/\sigma_B), \quad (4)$$

$$CNR = \frac{1}{n}\sum_{i=1}^{n} 10\log\left(|\mu_i - \mu_B|/\sqrt{\sigma_i^2 - \sigma_B^2}\right), \quad (5)$$

$$VAR = \sum_{j}\sum_{i} |I_{i,j} - \mu_I|, \quad (6)$$

where $I_{i,j}$ represents the pixel value at the i-th row and j-th column of the denoise image with $I_{max}$ the maximum pixel value of the denoised image and $\mu_I$ the mean of the whole image. The $\mu_i$ and $\sigma_i^2$ denote the mean and the variance of the i-th selected structure region while the $\mu_B$ and $\sigma_B^2$ represent the mean and the variance of the background region. Higher SNR and CNR values, along with lower VAR values, indicate higher quality of the images.

The comparison was conducted on 30 randomly selected images from the volumes that were previously unseen during training. Three regions of interest in RGC axon, IPL, and outer retina (light green box in Fig. 2) and one background region (orange box in Fig. 2) were used for metric calculations. The results were summarized in Table. 1. Consistent with visual inspection, S2F significantly outperformed other methods in all metrics (marked with bold font in Tab. 1), indicating significant enhancement of image contrast while maintaining low noise levels. Additionally, S2F also achieved the fastest convergence speed to obtain the results.

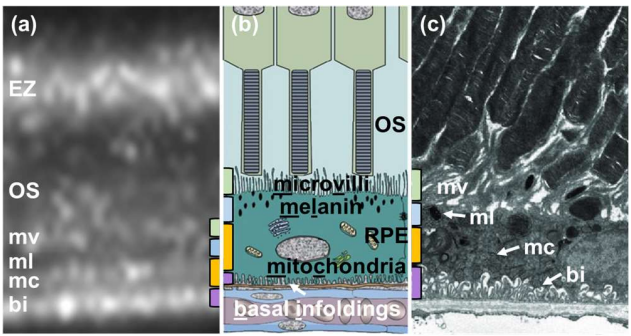

Fig. 3. (a) The zoomed outer retina region in the rat retinal image denoised by S2F. (b) Diagram of the outer retina. (c) Histology image of rat retina.

Table 1. Quantitative comparison of all models

|  | SNR | CNR | VAR (1e10) | Training speed |
|---|---|---|---|---|
| R1 | 32.55 | 1.61 | 5.67 | - |
| Merged | 35.55 | 2.24 | 4.84 | - |
| S2F | **58.43** | **5.44** | **2.98** | **25 epochs** |
| N2N | 53.78 | 3.67 | 3.87 | 101 epochs |
| N2V | 51.05 | 4.91 | 3.58 | 56 epochs |

Furthermore, we varied Gaussian window bandwidth to observe the effect on S2F performance. We conducted the comparison on five different bandwidths (10%, 25%, 50%, 75%, and 90% percentage of the raw spectrum) (Fig 4). More noticeable graininess and lower layer contrast were exhibited in the denoised images with larger bandwidth, which can be anticipated because the larger bandwidth, the higher structural similarity of the generated LR image to the full-spectrum HR image, thus failed in achieving the 1:N mapping with poorer denoising performance. With 90% or larger bandwidth, the S2F was almost deteriorated to the traditional N2N method. However, a shorter bandwidth (such as 10% bandwidth) also caused negative effects for the denoising. For example, oversmoothed and unrealistic RGC axon bundles (dashed box 1) and axial banding artifact (dashed box 2) emerged. In addition, the shadow of blood vessel (dashed box 3) almost disappeared in the denoised image. This phenomenon can be attributed to the over degradation of the resolution as the most important fine structures were lost in the sub-band image and the corresponding explanations may deviate from the real image. Nevertheless, the sublaminar structure of outer retina (dashed box 4), as a kind of high-frequency feature, was well maintained with clear and continuous visualization.

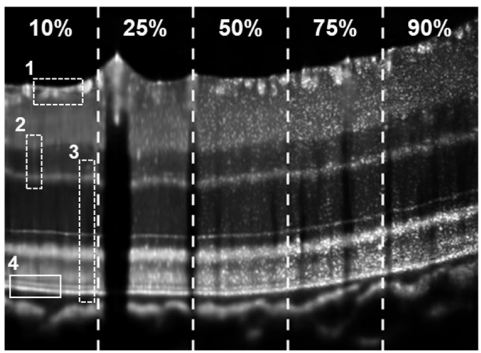

Fig. 4. Comparisons of S2F denoising performance with window bandwidth of 10%, 25%, 50%, 75% and 90% of the raw spectrum.

Initiated by the fast convergence and self-supervise feature of S2F, we tested its performance in a unique pipeline, i.e., using limited frames (10%, N=50) from a single volume to train and apply the dedicated model within that volume. Surprisingly, the performance did not deteriorate (SNR: 54.37 vs 54.39, CNR: 5.49 vs 5.34, VAR(1e10): 3.58 vs 3.49), and even better from visual inspection, when compared to that by the generalized model. Additionally, the training took less than 1 min to converge, making it practical for real-time applications to be integrated with OCT processing modules and be adaptive to various OCT scans quality in the future.

In summary, a self-supervised OCT despeckling method named Sub2Full was proposed. This approach works by utilizing the 1:N mapping of LR images to HR images from two independent observations. The S2F strategy outperformed two typical self-supervised schemes, N2N and blind-spot-based N2V, in denoising and preserving structures. Although validated with U-net and N2N in vis-OCT, the S2F strategy should also work in other network architectures, deep learning schemes, and OCT systems. We envision S2F as a promising solution for effective and practical OCT denoising to unlock new features and hope to apply it for commercialization and clinical utilization in the near future.

**Acknowledgements.** We appreciate the funding from the Knight Templar Eye Foundation, Alcon Research Institute, and Eye & Ear Foundation of Pittsburgh to Dr. Shaohua Pi and Dr. Bingjie Wang, as well as NIH CORE Grant P30 EY08098 and an unrestricted grant from Research to Prevent Blindness to the Department of Ophthalmology. We gratefully acknowledge the insightful discussion with Dr. Yuanyuan Chen.

**Disclosure.** The authors declare no conflicts of interest.